\documentclass[%
 reprint,
 amsmath,amssymb,
 aps,
]{revtex4-2}
\usepackage{lipsum}
\usepackage{graphicx}
\usepackage{dcolumn}
\usepackage{bm}
\usepackage{color}
\usepackage{comment}
\usepackage{physics}

\begin{document}
\preprint{APS/123-QED}
\title{Flat bands, quantum Hall effect and superconductivity in twisted bilayer graphene at magic angles}
\author{Leonardo A. Navarro-Labastida and Gerardo G. Naumis}
\date{December 2022}
\email{naumis@fisica.unam.mx}

\affiliation{%
Depto. de Sistemas Complejos, Instituto de F\'isica, Universidad Nacional Aut\'onoma de M\'exico (UNAM)\\
Apdo. Postal 20-364, 01000, Coyoac\'an, CDMX, M\'exico.
}%
\begin{abstract}
Flat band electronic modes are responsible for superconductivity in twisted bilayer graphene (TBG) rotated at magic angles. From there other magic angles can be found for any multilayered twisted graphene systems. Eventually this lead to the discovery of the highest ever known electron-electron correlated material. Moreover, the quantum phase diagram of TBG is akin to those observed among high-$T_{c}$ superconductors and thus there is a huge research effort to understand TBG in the hope of clarifying the physics behind such strong correlations. A particularity of the TBG is the 
coexistence of superconductivity and the fractional Quantum Hall effect, yet this relationship is not understood. In this work,  a simple $2\times 2$ matrix model for TBG is introduced. It contains the magic angles and  due to the intrinsic chiral symmetry in TBG, a lowest energy level related to the quantum Hall effect. The non-Abelian properties of this Hamiltonian play a central role in the electronic localization to produce the flat bands and here it is proved that the squared Hamiltonian of the chiral TBG model is equivalent to a single electron Hamiltonian inside of a non-Abelian pseudo-magnetic field produced by electrons in other layers. Therefore, the basic and fundamental elements in the physics of magic angles are determined. In particular, an study is made on these fundamental energy contributions at the $\Gamma$-point due to its relation to the recurrence of magic angles and its relationship with the Quantum Hall effect. 
\end{abstract}

\maketitle
\section{Introduction}

\begin{figure}[h!]
\includegraphics[scale=0.6]{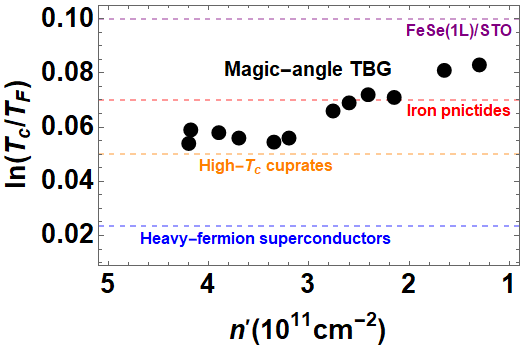}
\caption{Variation of $\ln{(T_c/T_F)}$ (logarithmic scale) as function of $n^{\prime}$ (charge doping) in scaled units of $10^{11} cm^{-2}$ for magic-angle TBG (blue points). Here $T_c$ is the critical temperature for a superconductor state and $T_F$ is the Fermi temperature. The horizontal lines indicate the approximate $T_c/T_F$ values of the corresponding family of materials. Adapted from Reference \cite{Cao2018}.}
\label{fig:TFTC}
\end{figure}

Recently it has been showed that twisted bilayer graphene systems support superconducting phases at certain special twist angles \cite{Cao2018,Yankowitz2019}, where electronic correlations are maximized due to the existence of flat-bands \cite{Kerelsky2019}. Moreover, Cao et.al. \cite{Cao2018} found that such TBG systems have a Mott insulating phase that appears in the middle of unconventional superconducting phases, similar to the phase diagram found in cuprates and other high temperature superconductors \cite{Fradkin2015}. 
The appearance of a Mott correlated insulator and unconventional superconducting phases in the flat band of magic-angle TBG at a small carrier density cannot be explained by weak-coupling BCS theory \cite{2005Andreev,2015fRADKINN,Fidrysiak2018,Roy2019,2019Sarma,2020Yuan,Khalaf2020,kHALAF_B2020,Ledwith2021,2021Tanigushi}. In fact, although $T_c$ is very small, of the order of $1.7K$, the electron-electron coupling turns out to be very high. In Fig. \ref{fig:TFTC} we show the variation of $\ln{(T_c/T_F)}$, where $T_F$ is the Fermi temperature and $T_c$ is the critical temperature of the superconducting phase as a function of the charge doping $n^{\prime}$.
This plot intends to compare the electron-electron coupling once the density of carriers and $T_F$ are taken into account. The magic-angle TBG $T_c/T_F$ ratio is above the trend lines on which most heavy fermions, cuprates and organic superconductors lie \cite{Cao2018}. Thereafter, it has been experimentally found that trilayer graphene rotated by magic angles turns out to be the highest ever found electron-electron correlated material \cite{2020Jeong}. Up to now, the origin of superconductivity in TBG remains under debate.

A interesting strong coupling theory of superconductivity based on skyrmions could explain the mechanism of superconductivity in TBG \cite{Khalaf2020}. For this purpose is important to understand the quantum geometry of flat bands in magic-angle TBG and the interplay between interaction range and Berry curvature inhomogeneity. A deeper theoretical study by Ledwith et.al. \cite{2022ledwithv} based in vortexability and the importance of chiral symmetry to induce vortexable bands as a generalization of the LLL (Lowest Landau Level) for hosting a short-range interacting ground state (SRI-GS) is of great relevance. This treatment of vortexable bands allows the construction of exact many-body FQHE (Fractional Quantum Hall Effect) ground states in the limit of short-range interactions, and possibly vortexable bands of equal and opposite Chern numbers related to superconductors based on skyrmions. On the other hand, magic-angle TBG has flat Chern bands at zero magnetic field. Therefore, TBG promises a route towards stabilizing zero-fields FCIs \cite{2021jarillo}. 

In fact, the study of the electronic properties of twisted bilayer graphene started before the discovery of superconductivity at magic angles. In works of 2007 by J. Santos \cite{Santos2007} and 2011 by A. Macdonald \cite{MacDonald2011}, an effective low energy continuous Hamiltonian model was derived. 

\begin{figure}[h!]
\includegraphics[scale=0.5]{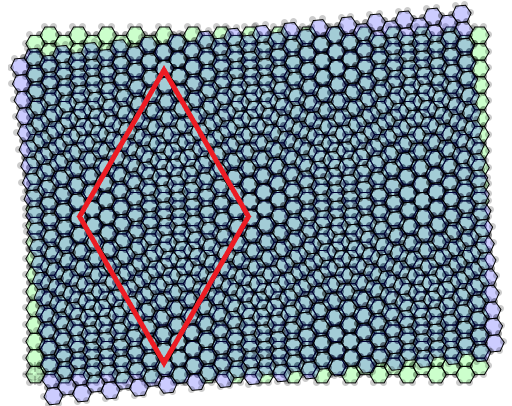}
\caption{Picture of twisted bilayer graphene (TBG) system. The graphene monolayer 1 (green) and monolayer 2 (purple) have a relative twist angle that produce a Moiré pattern (cyan) with a unit cell much bigger than the original unit cell of each monolayer graphene. The unit cell of the Moiré pattern is indicated (red).}
\label{fig:unitcell}
\end{figure}

In such studies the idea was to generate Moire patterns as a function of the twist angle between graphene layers. In Fig. \ref{fig:unitcell} we show the unit cell for the Moiré pattern of TBG. The new unit cell vectors in the Moiré lattice are scaled as the inverse of the twist angle between layers, therefore, for small twist angles, it is expected to have a bigger Moiré unit cell. Form there is possible to define a Moire Brillouin zone (mBZ). For small twist angles, the mBZ is also small due to the Moiré modulation vector. Flat-band were found at certain angles and from there superconductivity was predicted  \cite{MacDonald2011}.

The reason for the appearance of flat bands and its quantized nature is not understood. Therefore, lots of research have been conducted in this direction. Notice that flat bands can also occur in graphene over substrates even without twists yet TBG systems are paradigmatic. To make the TBG model more realistic, G. Tarnopolsky et. al. \cite{Tarnpolsky2019} took into account the structural relaxation due to carbon-carbon repulsion between layers and a chiral continuum model was produced.  Perhaps, so far, it is the simplest and more realistic model that best captures the nature of magic angles; at these special angles the dispersion energy at the lowest bands becomes flat and has a recurrence behavior. Also at these magic angles the Fermi velocity goes to zero.
Due to its chiral symmetry, the Hamiltonian of this model produces an intra-valley inversion symmetry \cite{Wang2020} so the energy dispersion is inversion symmetric at all twist angles and thus symmetry protected at any twist angle. The zero-mode have some resemblance to the ground state of a quantum Hall effect wave function on a torus \cite{Tarnpolsky2019,Popov2020}, and, therefore, the solution is of the quantum harmonic oscillator type, where Landau levels arise \cite{Hejazi2019,Uri2020,Benlakhouy2022}. Another interesting mathematical characteristic is that flat-band modes are constructed from a complex analytic functions that are ratios of the theta Jacobi functions \cite{Tarnpolsky2019,Popov2020}.\\
Furthermore, due to the quantized nature of the magic angles it appears that there is an adiabatic change in integers of this topological invariant. 
However, there are many open question concerning the possible relationship between the quantum hall effect (QHE) and the TBG, the quantized nature of flat-bands and the nature of wave functions. 

In a series of previous papers, the authors showed that by making a supersymmetric transformation, the physics behind the magic angles becomes clear \cite{Naumis2021,Naumis2022,2022NN}. The aim of this work is to review the topic and clarify the physical behavior of this model as well as to develop an effective equation for the Hamiltonian in order to study all other non flat band states. Although one of the authors presented before in this journal some details of this model in a review concerning 2D materials \cite{Naumis_2021}, here we go further by showing that the effective system is related to pseudo-magnetic fields, bringing the possibility of writing the Hamiltonian in a non-Abelian fashion where effective magnetic fields appear, making the connection with the QHE transparent.  Then we discuss the physical picture that arises from this minimal model including an analytical calculation of the first magic angle to show how this arises as a balance between kinetic, confinement energies and interlayer currents.

\section{Chiral TBG system}
The Bistritzer-MacDonald Hamiltonian (BMH) model arises by a simple stacking of two rotated graphene layers with an interaction between them \cite{MacDonald2011}. One of the authors gave a simple derivation of this model in this journal before and thus we refer the reader to it \cite{Naumis_2021}. Although the BMH contains the magic angle physics, in real TBG stacking points where carbon atoms are on top of another carbon (known as AA stacking) are pulled apart out of the intralayer plane due to Coulomb repulsion. As a consequence, interlayer electron tunnelling in AA sectors is greatly reduced and can be safely neglected.  This result in a chiral Hamiltonian (CH)  \cite{Tarnpolsky2019,Ledwith2021}. Let us write this model. Consider as basis the wave vectors  $\Phi(r)=\begin{pmatrix} 
\psi_1(r) ,
\psi_2(r),
\chi_1(r),
\chi_2(r)
\end{pmatrix}^T$ where the index $1,2$ represents each graphene layer and $\psi_j(r)$ and $\chi_j(r)$ are the Wannier orbitals on each site of the graphene's unit cell, i.e., in sites A and B of the graphene's bipartite lattice.  Using this base, the CH  is \cite{Tarnpolsky2019,Khalaff2019, Ledwidth2020}, 
\begin{equation}
\begin{split}
\mathcal{H}
&=\begin{pmatrix} 
0 & D^{\ast}(-r)\\
 D(r) & 0
  \end{pmatrix}  \\
\end{split} 
\label{H_initial}
\end{equation}
where we defined the zero mode operator as, 
\begin{equation}
\begin{split}
D(r)&=\begin{pmatrix} 
-i\Bar{\partial} & \alpha U(r)\\
  \alpha U(-r) & -i\Bar{\partial} 
  \end{pmatrix}  \\
\end{split} 
\end{equation}
and its rotated and conjugated version,
\begin{equation}
\begin{split}
D^{*}(-r)&=\begin{pmatrix} 
-i\partial & \alpha U^{*}(-r)\\
  \alpha U^{*}(r) & -i\partial 
  \end{pmatrix}  \\
\end{split} 
\end{equation}
We also defined the antiholonomic operator,
\begin{equation}
\Bar{\partial}=\partial_x+i\partial_y
\end{equation}
and the holonomic operator,
\begin{equation}
    \partial=\partial_x-i\partial_y,
\end{equation}
Both operators play an important role in the theory. 
The interlayer potential is \cite{Tarnpolsky2019},
{\footnotesize
\begin{equation}
    U(\bm{r})=e^{-i\bm{q}_1\cdot \bm{r}}+e^{i\phi}e^{-i\bm{q}_2\cdot \bm{r}}+e^{-i\phi}e^{-i\bm{q}_3\cdot \boldsymbol{r}}
\end{equation}
}
Here $\phi=2\pi/3$ and $\bm{q}_{1}=k_{\theta}(0,-1)$, $\bm{q}_{2}=k_{\theta}(\frac{\sqrt{3}}{2},\frac{1}{2})$, $\bm{q}_{3}=k_{\theta}(-\frac{\sqrt{3}}{2},\frac{1}{2})$.\\

The distance between each layer graphene Dirac cone is $k_{\theta}=2k_{D}\sin{\frac{\theta}{2}}$ with $k_{D}=\frac{4\pi}{3a_{0}}$ the Dirac wave vector and $a_{0}$ the lattice constant of graphene. The physics of this model is captured via the parameter $\alpha$, defined as,
\begin{equation}
    \alpha=\frac{w_1}{v_0 k_\theta}
\end{equation}
where $w_1$ is the interlayer coupling of stacking AB/BA with value $w_1=110$ meV and $v_0$ is the Fermi velocity with value $v_0=\frac{19.81eV}{2k_D}$.\\

Here we will use units where $v_0=1$, $k_{\theta}=1$, thus the antiholonomic and holonomic operators are dimensionless and we can consider the system as if it were with a fixed geometry, i.e., setting $\bm{q}_{1}=(0,-1)$, $\bm{q}_{2}=(\frac{\sqrt{3}}{2},\frac{1}{2})$, $\bm{q}_{3}=(-\frac{\sqrt{3}}{2},\frac{1}{2})$ while the twist angle enters only in the coupling parameter $\alpha$.  The vectors $\bm{b}_{1,2}=\bm{q}_{2,3}-\bm{q}_{1}$ turn out to be the Moir\'e reciprocal vectors and for its utility, we define a third vector $\bm{b}_{3}=\bm{q}_{3}-\bm{q}_2$.

To finish the model, we found useful to define a set of unitary vectors $\bm{\hat{q}}_{\mu}^{\perp}$ given by
$\bm{\hat{q}}_{1}^{\perp}=(1,0),  \bm{\hat{q}}_{2}^{\perp}=\big(-\frac{1}{2},\frac{\sqrt{3}}{2}\big), 
    \bm{\hat{q}}_{3}^{\perp}=\big(-\frac{1}{2},-\frac{ \sqrt{3}}{2}\big)$.
    

 \section{Squared TBG Hamiltonian and Pseudo-magnetic fields}
\begin{figure}[h!]
\includegraphics[scale=0.25]{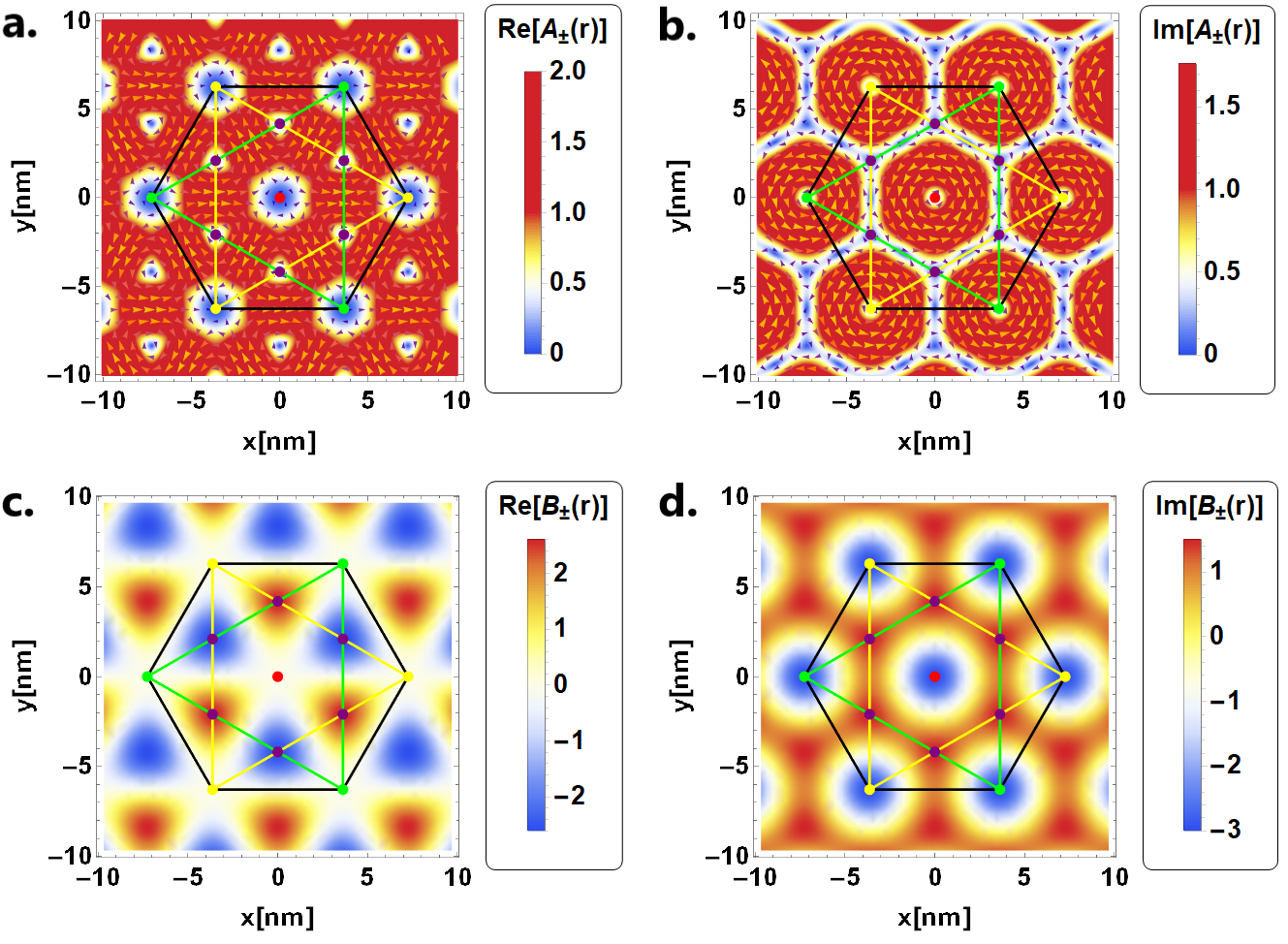}
\caption{In panels a) and b), we show the real and imaginary parts of the pseudo-magnetic vector potential $\bm{A}_{\mp}(\bm{r})$, respectively. In panels c) and d), we show the real and imaginary parts of the pseudo-magnetic scalar potential $B_{\mp}(\bm{r})$, respectively. The arrows denote the direction of the vectors while the color is used for the magnitude. In all panels we indicate the stacking points AA (red), BA (yellow) and AB (green) over the real-space unit cell (black hexagon). The purple dots are related to the centers of charge for $\alpha\rightarrow\infty$ in the electronic density, and are also related to the QHE \cite{2022NN}.}
\label{fig:AB_potentials}
\end{figure}

  In a previous work \cite{Naumis2021} we showed how, by taking the square of $H$, which is akin to consider a supersymmetric model, it is possible to write the Hamiltonian as a $2 \times 2$ matrix. This transformation takes into account the particle-hole symmetry and folds the band around $E=0$. Physically, it removes one of the bipartite lattices on each layer, leaving two triangular lattices on each layer. In a series of previous works we investigated several properties of this Hamiltonian, among the most important is the mapping of the flat-band into a ground state separated by a gap form the rest of the states. This state has an antibonding nature in a triangular lattice and then has frustration, associated with a massive degeneration \cite{Naumis1994}.  Other important recent articles exploring several properties such as topology and supersymmetry on lattice systems of the squared Hamiltonian are of great relevance and thus we refer the reader to them \cite{Hatsugai2020,2022Krishanu,2022Hatsugai,Hatsugai2021}.
For the present work we decided to change the notation used in such works and write $H^{2}$ in terms of a more physically suggestive picture by using pseudo-magnetic fields, 
\begin{equation}\label{eq:H2BA}
\begin{split}
H^{2}&=\begin{pmatrix} -\nabla^{2}+\alpha^{2}U_{-} &  -i2\alpha\bm{A}_{-}\cdot\nabla+\alpha B_{-} \\
 -i2\alpha\bm{A}_{+}\cdot\nabla+\alpha B_{+} & -\nabla^{2}+\alpha^{2}U_{+}
  \end{pmatrix} 
  \end{split} 
  \end{equation}
where  $U_{\pm}$ denotes the intralayer confinement  potentials \cite{Naumis2021},
\begin{equation}\label{eq:UUmasmenos}
\begin{split}
U_{\mp}=|U(\mp\bm{r})|^{2}
  \end{split}. 
  \end{equation}
 and, 
\begin{equation}\label{eq:AAvector}
\begin{split}
\bm{A}_{\pm}(\bm{r})= \sum_{\mu}e^{\pm i\bm{q}_{\mu}\cdot\bm{r}}\hat{\bm{q}}^{\perp}_{\mu}
\end{split} 
  \end{equation}
which turns out to be a two-dimensional pseudo-magnetic vector potential in the Coulomb gauge as is easy to prove that $\nabla\cdot\bm{A}_{\pm}=0$. We also define a pseudo-magnetic scalar potential,
\begin{equation}\label{eq:BBScalar}
\begin{split}
B_{\pm}(\bm{r}) \equiv  \mp i\sum_{\mu}e^{\pm i\bm{q}_{\mu}\cdot\bm{r}}
  \end{split} 
  \end{equation}
which has the property, 
\begin{equation}
    B_{\mp}(\bm{r})=\nabla \times \bm{A}_{\mp}(\bm{r})
\end{equation} 
where the cross product between two-dimensional vectors given by $\bm{A}=(A_x,A_y)$ and $\bm{B}=(B_x,B_y)$ is $\bm{A} \times \bm{B}=A_x B_y-A_yB_x$.  Observe that the related pseudo-magnetic flux over the real-space unit cell vanishes,
\begin{equation}
\begin{split}
\Phi_{\pm}=\int{B_{\pm}(\bm{r})\cdot d^{2}\bm{r}}=0
  \end{split} 
  \end{equation}
Such cancelling can be graphically understood in Figs. \ref{fig:AB_potentials}(a)-(b) where we plot the pseudo-magnetic vector potential in real space.
On the other hand, in Figs. \ref{fig:AB_potentials}(c)-(d) is shown the pseudo-magnetic scalar field. In both cases, we have a Kagome geometry for the pseudo-magnetic fields of the squared hamiltonian.

\section{Non-Abelian model for TBG}

Let us now explore the nature of eq. (\ref{eq:H2fundamental}). In analogy to an electron in a pseudo-magnetic vector potential, we define the canonical momentum,
\begin{equation}
\begin{split}
\bm{\Tilde{\Pi}}&=\bm{p}+\frac{e\alpha}{c}\hat{\bm{A}}
  \end{split} 
  \end{equation}
where $\bm{\Tilde{\Pi}}$ is the canonical $SU(2)$ momentum, $e$ the electron charge and $\hat{\bm{A}}$ is a non-Abelian $SU(2)$ (see below) pseudo-magnetic vector potential,
\begin{equation}
\begin{split}\label{eq:Non-abelian_pseudo}
\hat{\bm{A}}&=(\hat{\bm{A}}_{x},\hat{\bm{A}}_{y})
  \end{split} 
  \end{equation}
with $\hat{\bm{A}}_{x}=\bm{A}_{1,x}\hat{\tau_1}+\bm{A}_{2,x}\hat{\tau}_2$ and $\hat{\bm{A}}_{y}=\bm{A}_{1,y}\hat{\tau}_1+\bm{A}_{2,y}\hat{\tau}_2$, where we used the set of Pauli  matrices $\hat{\tau}_j$ (with $j=1,2,3$) in the pseudo-spin of the layer degree, and the $2 \times 2$ identity matrix $\hat{\tau}_0$.  Explicitly, the components of $\hat{\bm{A}}$ are,
\begin{equation}
\begin{split}
\bm{A}_{1,x}&=\cos{(\bm{q}_{\mu}\cdot\bm{r})}\bm{q}_{\mu}^{\perp,x},\\
\bm{A}_{1,y}&=\cos{(\bm{q}_{\mu}\cdot\bm{r})}\bm{q}_{\mu}^{\perp,y},\\
\bm{A}_{2,x}&=\sin{(\bm{q}_{\mu}\cdot\bm{r})}\bm{q}_{\mu}^{\perp,x},\\
\bm{A}_{2,y}&=\sin{(\bm{q}_{\mu}\cdot\bm{r})}\bm{q}_{\mu}^{\perp,y}.
  \end{split} 
  \end{equation}

Note that $\hat{\bm{A}}$ is non-Abelian as follows from the fact that $[\hat{\bm{A}}_{\mu}(\bm{r}),\hat{\bm{A}}_{\nu}(\bm{r}^{\prime})]\neq 0$, i.e.,  it does not commute with itself at different locations. 
  
Using the canonical momentum, the squared Hamiltonian  can be written as, 
\begin{equation}
\begin{split}\label{eq:square_momentum}
H^{2}&=(\bm{\Tilde{\Pi}}\cdot\bm{\Tilde{\Pi}})-i(\bm{\Tilde{\Pi}}\times\bm{\Tilde{\Pi}})\\
&=(-i\bm{\nabla}+\alpha\hat{\bm{A}})^{2}-i\epsilon_{ijk}[\bm{\nabla}_{i}+\alpha\hat{\bm{A}}_{i},\bm{\nabla}_{j}+\alpha\hat{\bm{A}}_{j}]
  \end{split} 
  \end{equation}
Eq. (\ref{eq:square_momentum}) can be further simplified to, 
\begin{equation}
\begin{split}\label{eq:square_tensorform}
H^{2}&=-\nabla^{2}+\alpha^{2}\hat{\bm{A}}^{2}-2i\alpha(\hat{\bm{A}}\cdot\nabla)+\alpha\hat{\bm{F}}_{\mu\nu}
  \end{split} 
  \end{equation}
where $\hat{\bm{F}}_{\mu\nu}=\partial_{\mu}\hat{\bm{A}}_{\nu}-\partial_{\nu}\hat{\bm{A}}_{\mu}-i\alpha[\hat{\bm{A}}_{\mu},\hat{\bm{A}}_{\nu}]$ is the Zeeman coupling term with natural units $\hbar=m=e=1$.   
Therefore, eq. (\ref{eq:square_tensorform}) proves that indeed the squared chiral Hamiltonian describes an electron in a non-Abelian $SU(2)$ pseudo-magnetic field as was suggested before \cite{Guinea2012}. 

Now we investigate the Zeeman term. We start by observing that,
\begin{equation}
\begin{split}\label{eq:A_conmutator}
-i[\hat{\bm{A}}_{x}(\bm{r}),\hat{\bm{A}}_{y}(\bm{r}^{\prime})]&=-i[\bm{A}_{1x}\hat{\tau}_1+\bm{A}_{2x}\hat{\tau}_2,\bm{A}_{1y}\hat{\tau}_1+\bm{A}_{2y}\hat{\tau}_2]\\
&=-i[(\bm{A}_{1x}\bm{A}_{2y}-\bm{A}_{1y}\bm{A}_{2x})\hat{\tau}_1\hat{\tau}_2\\
&+(\bm{A}_{2x}\bm{A}_{1y}-\bm{A}_{2y}\bm{A}_{1x})\hat{\tau}_2\hat{\tau}_1]
\end{split} 
\end{equation}
from where, 
\begin{equation}
\begin{split}\label{eq:Axy}
\bm{A}_{1x}\bm{A}_{2y}-\bm{A}_{1y}\bm{A}_{2x}=\sum_{\mu,\nu}\sin{(\bm{q}_{\nu}\cdot\bm{r}^{\prime}-\bm{q}_{\mu}\cdot\bm{r})}\hat{q}_{\mu}^{\perp x}\hat{q}_{\nu}^{\perp y}
\end{split}
\end{equation}
Substituting Eq.(\ref{eq:Axy}) in the commutator Eq.(\ref{eq:A_conmutator}), it follows that for $\bm{r}=\bm{r}^{\prime}$ , 
\begin{equation}
\begin{split}\label{eq:A_conm2}
-i[\hat{\bm{A}}_{x}(\bm{r}),\hat{\bm{A}}_{y}(\bm{r})]&=2\hat{\tau}_{3}\sum_{\mu,\nu}\sin{([\bm{q}_{\nu}-\bm{q}_{\mu}]\cdot\bm{r})}\hat{q}_{\mu}^{\perp x}\hat{q}_{\nu}^{\perp y}
\end{split}
\end{equation}
where we specified that the conmutator must be taken at the same point. We also used that $2i\hat{\tau_{3}}=[\hat{\tau}_{1}, \hat{\tau}_{2} ]$. From Eq.(\ref{eq:A_conm2}) it follows that, 
\begin{equation}
\begin{split}\label{eq:Axy2}
&2\sum_{\mu,\nu}\sin{([\bm{q}_{\nu}-\bm{q}_{\mu}]\cdot\bm{r})}\hat{q}_{\mu}^{\perp x}\hat{q}_{\nu}^{\perp y}=2[\sin{(\bm{b}_1\cdot\bm{r})}\frac{\sqrt{3}}{2}\\
&-\sin{(\bm{b}_2\cdot\bm{r})}\frac{\sqrt{3}}{2}+\sin{(\bm{b}_3\cdot\bm{r})}\frac{\sqrt{3}}{2}]\\
&=\sqrt{3}\sum_{\mu}(-1)^{\mu-1}\sin{(\bm{b}_{\mu}\cdot\bm{r})}
\end{split}
\end{equation}
multiplying by $\alpha^{2}$ Eq.(\ref{eq:Axy2}) and substituting in  Eq.(\ref{eq:A_conm2}),
\begin{equation}
\begin{split}\label{eq:A_conm3}
-i\alpha^{2}[\hat{\bm{A}}_{x}(\bm{r}),\hat{\bm{A}}_{y}(\bm{r})]&=\sqrt{3}\alpha^{2}\hat{\tau}_{3}\sum_{\mu}(-1)^{\mu-1}\sin{(\bm{b}_{\mu}\cdot\bm{r})}\\
& \equiv h_{3}(\bm{r})\hat{\tau}_{3}
\end{split}
\end{equation}
where we defined the potential of the non-Abelian term  $h_{3}(\bm{r})=\sqrt{3}\alpha^{2}\sum_{\mu}(-1)^{\mu-1}\sin{(\bm{b}_{\mu}\cdot\bm{r})}=\frac{\alpha^{2}}{2}(U_{-}-U_{+})$. \\
\\
\begin{figure}[h!]
\includegraphics[scale=0.25]{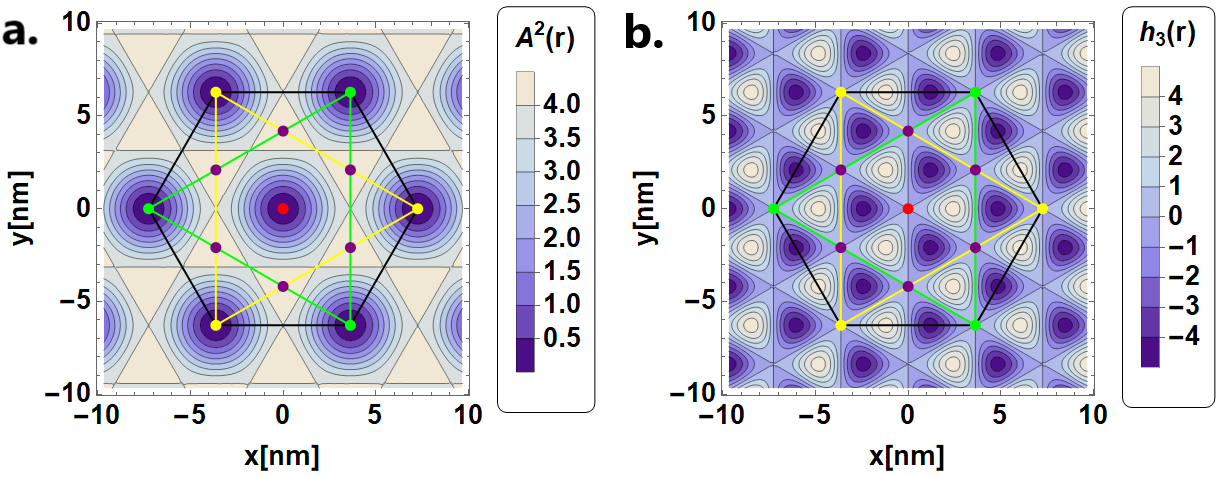}
\caption{Representation in real space of a) $\bm{A}^{2}(\bm{r})$ and b) $\bm{h}_{3}(\bm{r})$. Where is indicated the stacking points AA (red), BA (yellow) and AB (green) over the real-space unit cell (black hexagon). The purple dots are related to the centers of charge for $\alpha\rightarrow\infty$ in the electronic density, and are also related to the QHE \cite{2022NN}.}\label{fig:AAhz}
\end{figure}

In Fig.\ref{fig:AAhz} we show the contour plot in real space for the potential  $h_{3}(\bm{r})$. As in Fig. \ref{fig:AB_potentials}, it can be seen that the pseudo-magnetic fields of the squared Hamiltonian behave like a Kagome lattice. Moreover, note that these pseudo-magnetic fields are described in a larger lattice due to the renormalization process of the graphene lattice into a triangular lattice. Therefore, the original real-space unit cell is now three times larger where the pseudo-magnetic fields operate. \\

The other term of the Zeeman coupling is related to the pseudo-magnetic scalar potentials as follows, 
\begin{equation}
\begin{split}\label{eq:B_1}
\partial_{\mu}\hat{\bm{A}}_{\nu}-\partial_{\nu}\hat{\bm{A}}_{\mu}&=\sum_{\mu}[\sin{(\bm{q}_{\mu}\cdot\bm{r})}(-\hat{\bm{q}}_{\mu}^{\perp y}\bm{q}^{x}_{\mu}+\hat{\bm{q}}_{\mu}^{\perp x}\bm{q}^{y}_{\mu})\hat{\tau}_1 \\
&+\cos{(\bm{q}_{\mu}\cdot\bm{r})}(\hat{\bm{q}}_{\mu}^{\perp y}\bm{q}^{x}_{\mu}-\hat{\bm{q}}_{\mu}^{\perp x}\bm{q}^{y}_{\mu})\hat{\tau}_2]
\end{split}
\end{equation}
but $\bm{q}_{\mu}\times\hat{\bm{q}}^{\perp}_{\mu}=\hat{\bm{q}}_{\mu}^{\perp y}\bm{q}^{x}_{\mu}-\hat{\bm{q}}_{\mu}^{\perp x}\bm{q}^{y}_{\mu}=1$, therefore Eq. (\ref{eq:B_1}), 
\begin{equation}
\begin{split}\label{eq:B_2}
\partial_{\mu}\hat{\bm{A}}_{\nu}-\partial_{\nu}\hat{\bm{A}}_{\mu}&=\sum_{\mu}[\sin{(\bm{q}_{\mu}\cdot\bm{r})}\hat{\tau}_1-\cos{(\bm{q}_{\mu}\cdot\bm{r})}\hat{\tau}_2]\\
&=\begin{pmatrix}
0 & i\sum_{\mu}e^{-i\bm{q}_{\mu}\cdot\bm{r}}\\
-i\sum_{\mu}e^{i\bm{q}_{\mu}\cdot\bm{r}}& 0
  \end{pmatrix} \\
&=\begin{pmatrix}
0 & B_{-}\\
B_{+} & 0
  \end{pmatrix}
\end{split}
\end{equation}
where $B_{\pm}(\bm{r})$ is the pseudo-magnetic  interlayer potential  and the non-Abelian pseudo-magnetic vector potential is written in matrix form as,
\begin{equation}
\begin{split}
\hat{\bm{A}}&=\sum_{\mu}[\cos{(\bm{q}_{\mu}\cdot\bm{r})}\hat{\tau}_1+\sin{(\bm{q}_{\mu}\cdot\bm{r})}\hat{\tau}_2]\bm{q}^{\perp}_{\mu}\\
&=\begin{pmatrix}
0 & \sum_{\mu}e^{-i\bm{q}_{\mu}\cdot\bm{r}}\bm{q}^{\perp}_{\mu}\\
\sum_{\mu}e^{i\bm{q}_{\mu}\cdot\bm{r}}\bm{q}^{\perp}_{\mu}& 0
  \end{pmatrix} \\
&=\begin{pmatrix}
0 & \bm{A}_{-}\\
\bm{A}_{+} & 0
 \end{pmatrix}
\end{split}
\end{equation}
In the same way, the square of the non-Abelian pseudo-magnetic vector potential $\hat{\bm{A}}$ is related to a new potential $V(\bm{r})=\alpha^{2}\bm{A}^{2}$ where, 
\begin{equation}
\begin{split}\label{eq:A2_1}
\hat{\bm{A}}^{2}&=(\bm{A}^{2}_{1x}+\bm{A}^{2}_{2x}+\bm{A}^{2}_{1y}+\bm{A}^{2}_{2y})\hat{\tau}_{0}\\
&=\sum_{\mu,\nu}\cos{([\bm{q}_{\mu}-\bm{q}_{\nu}]\cdot\bm{r})}\bm{q}^{\perp}_{\mu}\cdot\bm{q}^{\perp}_{\nu}\hat{\tau}_{0}\\
&=(3-\sum_{\mu}\cos{(\bm{b}_{\mu}\cdot\bm{r})})\hat{\tau}_{0}
  \end{split} 
  \end{equation}
It follows that,
\begin{equation}
    V(\bm{r})=\alpha^{2}(3-\sum_{\mu}\cos{(\bm{b}_{\mu}\cdot\bm{r})})=\frac{\alpha^{2}}{2}(U_{-}+U_{+})
\end{equation}
and therefore $\alpha^{2}\hat{\bm{A}}^{2}=V(\bm{r})\hat{\tau}_{0}$. Fig. \ref{fig:AAhz} presents $\bm{A}^2$ which in fact is a confinement potential akin to a quantum dot.

Using all these results, it is possible to write the square Hamiltonian Eq. (\ref{eq:H2BA}) in a way that manifest more clearly its non-Abelian nature. We start by writing $H^{2}$ purely in terms of $\bm{A}$,
\begin{equation}
\begin{split}
&H^{2}=\\
&\begin{pmatrix}
-\nabla^{2}+\alpha^{2}(\bm{A}^{2}-i[\bm{A}_{x},\bm{A}_y]) &  \alpha(-2i\bm{A}_{-}\cdot\nabla+\nabla \times \bm{A}_{-})\\
 \alpha(-2i\bm{A}_{+}\cdot\nabla+\nabla \times \bm{A}_{+}) & -\nabla^{2}+\alpha^{2}(\bm{A}^{2}+i[\bm{A}_{x},\bm{A}_y])
  \end{pmatrix} 
  \end{split} 
  \label{eq:H2fundamental}
  \end{equation}
where $[\bm{A}_{x},\bm{A}_y]$ denotes the commutator of operators $\bm{A}_{x}$ and $\bm{A}_y$.  Also we can write
$H^{2}$ as follows,
\begin{equation}
    \begin{split}
H^{2}=(-\nabla^{2}+V(\bm{r})) \hat{\tau_0}+h_3(\bm{r}) \hat{\tau}_3
+h_1(\bm{r})\hat{\tau}_1+h_2(\bm{r})\hat{\tau}_2
  \end{split} 
\end{equation}
where we defined $h_3(\bm{r})=-\alpha^{2}i[\bm{A}_{x}(\bm{r}),\bm{A}_y(\bm{r})]$ and $V(\bm{r})=\alpha^{2}\bm{A}^{2}$. Also we define the off-diagonal components as $h_1(\bm{r})=-\sum_{\mu}[\sin{(\bm{q}_{\mu}\cdot\bm{r})}+2i\cos{(\bm{q}\cdot\bm{r})}\hat{\bm{q}}^{\perp}_{\mu}\cdot\bm{\nabla}]$ and $h_2(\bm{r})=\sum_{\mu}[\cos{(\bm{q}_{\mu}\cdot\bm{r})}-2i\sin{(\bm{q}\cdot\bm{r})}\hat{\bm{q}}^{\perp}_{\mu}\cdot\bm{\nabla}]$
. This last expression is very important as it manifest in a clear fashion the topological nature of the hamiltonian as it includes all three Pauli matrices \cite{Naumis2021}.

As proved in Ref. \cite{Naumis_2021}, in the limit $\alpha\rightarrow\infty$ we have that $\partial_{\mu}\hat{\bm{A}}_{\nu}-\partial_{\nu}\hat{\bm{A}}_{\mu}\gg-i\alpha[\hat{\bm{A}}_{\mu},\hat{\bm{A}}_{\nu}]$. Thus the Zeeman coupling is $\hat{\bm{F}}_{\mu\nu}\approx\partial_{\mu}\hat{\bm{A}}_{\nu}-\partial_{\nu}\hat{\bm{A}}_{\mu}$. Therefore $h_3(\bm{r})$ is neglected and the system has a pure Abelian field for higher magic angles \cite{2022NN}. The pure Abelian field means that 
a constant effective magnetic can be found and eventually, it leads to the proof that in such limit, the squared TBG hamiltonian at magic angles and the Quantum Hall effect hamiltonian are the same \cite{2022NN}. The other limit $\alpha \rightarrow 0$ can be  explored using perturbation theory, as done in the following section.


 \section{Strong non-Abelian limit: first magic angle perturbative solution at $\Gamma$-point}

In this section we will explore the strong non-Abelian limit of the hamiltonian and its relationship with the first magic angle. Before doing so, and as explained elsewhere \cite{Naumis_2021}, the $\Gamma$ point can be used to reveal the magic angles as $E^2(\bm{\Gamma})$ is always the highest energy for the central band. 
This is seen in Fig. \ref{fig:EgHH}, where we plot
$E^2(\bm{\Gamma})$ as obtained from the numerical calculation detailed in the appendix. For a flat band to exist, the energy $E^{2}(\bm{\Gamma})$ must be zero.

\begin{figure}[h!]
\includegraphics[scale=0.5]{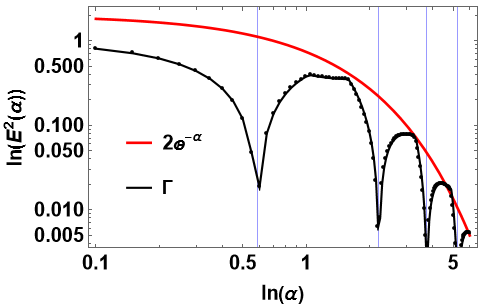}
\caption{$E^2(\bm{\Gamma})$ obtained from the squared Hamiltonian Eq. (\ref{eq:H2BA}) at the $\Gamma $-point where the vertical lines indicate the first four magic angles. The red curve represent the exponential squeezing of the bands for higher magic angles as was proved in Ref. \cite{Becker2020}.}
\label{fig:EgHH}
\end{figure}

The expected values for a given layer $j=1,2$ are, the kinetic energy,
\begin{equation}
\langle \bm{k} |\hat{T}_{j}|\bm{k}\rangle \equiv -\int_{m}\psi^{\dagger}_{\bm{k},j}(\bm{r})\nabla^{2}\psi_{\bm{k},j}(\bm{r})d^{2}\bm{r} 
\end{equation}
the confinement energy,
\begin{equation}\label{eq:Vdef}
   \langle \bm{k} |\hat{\bm{A}}^{2}_j|\bm{k} \rangle \equiv \alpha^{2}\int_{m}\bm{A}^{2}((\pm1)^{j}\bm{r}) \rho_{\bm{k},j}(\bm{r})d^{2}\bm{r}
\end{equation}
and the energy contribution from the off-diagonal operators,
\begin{equation}
\langle \bm{k}| \hat{\bm{A}}_j\cdot\nabla | \bm{k} \rangle \equiv -2i\alpha\int_{m}\psi^{\dagger}_{\bm{k},j+1}(\bm{r})(\bm{A}((\pm1)^{j}\bm{r})\cdot\nabla)\psi_{\bm{k},j}(\bm{r})d^{2}\bm{r}
\end{equation}

\begin{equation}
 \langle \bm{k}| \hat{B}_j | \bm{k}\rangle  \equiv \alpha\int_{m}\psi^{\dagger}_{\bm{k},j+1}(\bm{r}) B((\pm1)^{j}\bm{r})\psi_{\bm{k},j}(\bm{r})d^{2}\bm{r}
\end{equation}
here $\hat{\bm{A}}_j$ and $\hat{B}_j$ are the projections of the non-Abelian pseudo-magnetic vector potential and scalar potential, respectively, on each layer $j$ of the pseudo-spin degree. 
Therefore, the total energy for the squared Hamiltonian is, 
\begin{equation}
  \sum_j \langle \bm{k} |\hat{T} _{j}+\hat{\bm{A}}^{2}_{j}+\hat{\bm{A}}_j\cdot\nabla+\hat{B}_j|\bm{k}\rangle =E^2(k) 
\end{equation}
where the index $j=1,2$ for take into account the two layers.  

For the $\bm{\Gamma}$ point it is very illustrative to use  perturbation theory in the limit $\alpha \rightarrow 0$. The corresponding wave function is,
\begin{equation}\label{eq:psiatgamma}
\begin{split}
\psi_{\bm{\Gamma},1}(\bm{r})&=U(-\bm{r})+\frac{\alpha}{3} U(2\bm{r})+\frac{\alpha^{2}}{18}\left((2-e^{i\phi})U(-\sqrt{7}\bm{R}_{\gamma}\bm{r}) \right.\\ & \left.+(2-e^{-i\phi})U(-\sqrt{7}\bm{R}_{-\gamma}\bm{r})-4U(2\bm{r})\right)+...
\end{split}
\end{equation}
and $\psi_{\Gamma,2}(\bm{r})=i\mu_{\alpha}\psi_{\bm{\Gamma},1}(-\bm{r})$, where $R_{\gamma} \bm{r}$ is a counterclockwise rotation on angle $\gamma$ with $\tan(\gamma)= \sqrt{3}/5$ and $\mu_{\alpha}=\pm 1$, the minus sign is used for odd magic angles. In the $\Gamma$ point at a given layer we obtain up to second order in $\alpha$ that,

\begin{equation}
     \langle \bm{\Gamma} | \hat{T}  |\bm{\Gamma}  \rangle=\frac{1+\frac{4\alpha^2}{9}}{1+\frac{\alpha^2}{9}}, \hspace{0.3cm}   \langle \bm{\Gamma} | {\bm{\hat{A} }}^{2} |\bm{\Gamma}  \rangle=\frac{5\alpha^2}{1+\frac{\alpha^2}{9}}
\end{equation}

\begin{equation}
      \langle \bm{\Gamma} |  \hat{\bm{A}}\cdot\nabla |\bm{\Gamma}  \rangle=\frac{-3\alpha-\frac{2\alpha^2}{9}}{1+\frac{\alpha^2}{9}}, \hspace{0.3cm}  \langle \bm{\Gamma} | \hat{B} |\bm{\Gamma}  \rangle= \frac{-\alpha-\frac{11\alpha^2}{9}}{1+\frac{\alpha^2}{9}}
\end{equation}

It follows that,
\begin{equation}
  \langle \bm{\Gamma} |\hat{T} +{\bm{\hat{A} }}^{2}+ \hat{\bm{A}}\cdot\nabla+\hat{B}|\bm{\Gamma}\rangle =1-4\alpha+4\alpha^2\approx E^{2}(\bm{\Gamma})
  \label{eq:e2}
\end{equation}

\begin{figure}[h!]
\includegraphics[scale=0.7]{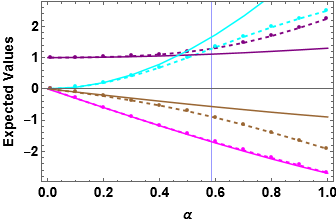}
\caption{Expected values in the $\bm{\Gamma}$ point vs $\alpha$. The numerical results are indicated with dashed lines with markers. The kinetic Energy $\langle\bm{\Gamma} | \hat{T}  |\bm{\Gamma}\rangle$ (purple), confinement energy $\langle\bm{\Gamma} |  {\bm{\hat{A} }}^{2}  |\bm{\Gamma}\rangle$ (cyan), $\langle\bm{\Gamma} | \hat{\bm{A}}\cdot\nabla |\bm{\Gamma}\rangle$ (magenta) and $\langle \bm{\Gamma} |\hat{B}|\bm{\Gamma}\rangle$ (brown). The solid lines are the perturbative solutions. The vertical line indicates the first magic angle value $\alpha_1=0.586$. Notice how    $\langle \bm{\Gamma} | \hat{T}  |\bm{\Gamma}\rangle=   \langle \bm{\Gamma} | {\bm{\hat{A} }}^{2}  |\bm{\Gamma} \rangle $ at $\alpha_1$. }
\label{fig:ExpectedSmallAlpha}
\end{figure}

In Fig. \ref{fig:ExpectedSmallAlpha} we compare these perturbative expected values with those obtained from the numerical results (see the appendix) showing a very good agreement for $\alpha \rightarrow 0$. 
Our perturbative analysis allows to get a glimpse of the first magic angle position as using eq. (\ref{eq:e2}) the condition for having a flat band turns out to be, 
\begin{equation}\label{eq:1stapprox}
    1-4\alpha+4\alpha^2=0 
\end{equation}

Therefore, $\alpha_1 \approx 1/2$, a value which is close to the numerically found first magic angle which is at $\alpha_1=0.586$. Higher order terms in the expansion are needed
to increase the accuracy, but yet the main principle behind a magic angle is already present in this simple approach.

Then we conclude  that in going from $\alpha=0$ to $\alpha_1$, the confinement potential starts to contribute and reaches the kinetic energy at the magic angle. The off-diagonal operators always diminish the energy.  As expected, the first magic angle is thus produced when the sum of the kinetic plus confinement energies are equal in magnitude to the expected values of the off-diagonal operators. For higher-order magic angles, we recently proved that the $H^2$ hamiltonian converges into the Quantum Hall effect hamiltonian, i.e., is equivalent to a two dimensional quantum oscillator \cite{2022NN}. Moreover, using boundary layer theory we were able to prove that flat-band states converge into coherent Landau levels opening a huge avenue of research not only experimentally but also technologically.

\section{Conclusion}
The chiral TBG Hamiltonian is equivalent to an electron inside of a non-Abelian pseudo-magnetic field. Here we gave its corresponding general expression. Then we explored the first magic angle using perturbation theory to show how the kinetic, confinement energy and interlayer currents produce a magic angle. For the first magic angle results that the non-Abelian properties are more important than other higher magic angles because the commutator becomes less important. Therefore, in the limit $\alpha\rightarrow\infty$, we recover an effective Abelian behavior that explains in part the effective equivalence between the Quantum Hall effect hamiltonian and the TBG hamiltonian \cite{2022NN}. The emergent Kagome structure observed in the pseudo-magnetic fields and for small angles $\alpha\rightarrow\infty$ \cite{2022NN,2018lado}, opens the question of whether there is any mechanism related to the Kagome lattice of the chiral magic angle TBG like spin-liquids or magnetic frustration, and what role it plays in the mechanism of superconductivity, FQHE and the physics of such strong correlations in TBG. This allows us to take advantage of this ideal scenario to build highly entangled ground states and opens many exciting technological possibilities in Moiré materials.

\section{acknowledgment}
This work was supported by UNAM DGAPA PAPIIT IN102620 (L.A.N.L. and G.G.N.) and CONACyT project 1564464. 

\section{Appendix}
\subsection{Exact diagonalization method for the squared Hamiltonian and its boundary conditions}
A general Bloch's wave function with momentum $\bm{k}$ in the mBZ has the form \cite{Tarnpolsky2019},
\begin{equation}
\begin{split}\label{eq:spinor2}
\Psi_{\bm{k}}(\bm{r})=\begin{pmatrix} \psi_{\bm{k},1}(\bm{r})\\
\psi_{\bm{k},2}(\bm{r})
  \end{pmatrix} =\sum_{m,n}\begin{pmatrix} a_{mn}\\
 b_{mn}e^{i\bm{q}_1\cdot\bm{r}}
  \end{pmatrix}e^{i(\bm{K}_{mn}+\bm{k})\cdot\bm{r}}
  \end{split} 
  \end{equation}
where $a_{mn}(b_{mn})$ are the Fourier coefficients of layer 1 (layer 2) and $\bm{K}_{mn}=m\bm{b}_1+n\bm{b}_2$ with $\bm{b}_{1,2}$ are the two Moiré Brillouin zone vectors and $m,n$ integers. If we plug Eq. (\ref{eq:spinor2}) in $H^2\Psi_{\bm{k}}(\bm{r})=E^2\Psi_{\bm{k}}(\bm{r})$ it follows that, 
  \begin{equation}
\begin{split}\label{eq:Eq1}
(|\bm{K}_{mn}+\bm{k}|^{2}+3\alpha_{0})a_{mn}+\alpha_{-\phi}a_{m+1,n}+\alpha_{\phi}a_{m,n+1}&\\
+\alpha_{\phi}a_{m-1,n}+\alpha_{-\phi}a_{m,n-1}+\alpha_{\phi}a_{m+1,n-1}+\alpha_{-\phi}a_{m-1,n+1}\\
+\alpha[(2\hat{\bm{q}}^{\perp}_1\cdot(\bm{K}_{mn}+\bm{k}+\bm{q}_1)-i)b_{mn}\\
+(2\hat{\bm{q}}^{\perp}_2\cdot(\bm{K}_{m+1,n}+\bm{k}+\bm{q}_1)-i)b_{m+1,n}\\
+(2\hat{\bm{q}}^{\perp}_3\cdot(\bm{K}_{m,n+1}+\bm{k}+\bm{q}_1)-i)b_{m,n+1}]=E^{2}a_{mn}
  \end{split} 
  \end{equation}
and
  \begin{equation}
\begin{split}\label{eq:Eq2}
(|\bm{K}_{mn}+\bm{k}+\bm{q}_1|^{2}+3\alpha_{0})b_{mn}+\alpha_{-\phi}b_{m-1,n}+\alpha_{\phi}b_{m,n-1}&\\
+\alpha_{\phi}b_{m+1,n}+\alpha_{-\phi}b_{m,n+1}+\alpha_{\phi}b_{m-1,n+1}+\alpha_{-\phi}b_{m+1,n-1}\\
+\alpha[(2\hat{\bm{q}}^{\perp}_1\cdot(\bm{K}_{mn}+\bm{k})+i)a_{mn}\\
+(2\hat{\bm{q}}^{\perp}_2\cdot(\bm{K}_{m-1,n}+\bm{k})+i)a_{m-1,n}\\
+(2\hat{\bm{q}}^{\perp}_3\cdot(\bm{K}_{m,n-1}+\bm{k})+i)a_{m,n-1}]=E^{2}b_{mn}
  \end{split} 
    \end{equation}
where $\hat{\bm{q}}^{\perp}_{\mu}=(\cos{[(\mu-1)\phi]}, \sin{[(\mu-1)\phi]})$ and $\alpha_{\phi}=e^{i\phi}\alpha^{2}$.
Here Eq. (\ref{eq:Eq1}) and Eq. (\ref{eq:Eq2}) form a coupled linear system that can be solved to find the corresponding eigenvalues. In general, there are $L=(2N+1)\times(2N+1)$ coefficients $a_{mn}$($b_{mn}$) with $N$ the range of the matrix and $2N+1$ the elements in the set, therefore, the Hamiltonian matrix has dimension $D=2L$.
The full space for the Fourier coefficients is spanned as, $S\rightarrow\big\{  \{m,-N,N\},\{n,-N,N\} \big\}$ where $S$ is the set of elements in each layer with range $2N+1$. \\
For simplicity, the geometry of the mBZ with $(m,n)$ coordinates is spanned as a square lattice and the immediately effect are the boundary conditions in each layer, 
\begin{equation}
\begin{split}
a_{\pm(N+1),n}=0\\
a_{m,\pm(N+1)}=0\\
b_{\pm(N+1),n}=0\\
b_{m,\pm(N+1)}=0
  \end{split} 
  \end{equation}
therefore, for $|M|>|N|$ all coefficients in both layers are zero. Also, symmetry considerations using the group $D_{3h}$ allow to simplify the system to optimize the numerical calculation. 
To find the expected values at the $\Gamma$ point, we need $\Psi_{\Gamma}(\bm{r})$, obtained by setting $\bm{k}=\bm{q}_1$.  

\bibliography{references.bib}
\end{document}